\documentclass[12pt, preprint, onecolumn, aps,longbibliography]{revtex4-1}
\usepackage{graphicx}

\usepackage{slashed}
\usepackage[small,justification=centerlast]{caption}
\usepackage{graphicx}
\graphicspath{ {images/} }
\usepackage{bm}
\usepackage{mathrsfs}
\usepackage[latin1]{inputenc}
\usepackage{array}
\usepackage{psfrag}
\usepackage{dsfont}
\usepackage{epsfig}
\usepackage{titletoc}
\usepackage{float}   
\usepackage{wrapfig} 
\usepackage{amsmath}
\usepackage[latin1]{inputenc}
\usepackage{amsmath}
\usepackage{amssymb}
\usepackage{cancel}
\usepackage{upgreek}
\usepackage{subfig}
\usepackage{tabularx}
\usepackage{color}
\usepackage{units}
\usepackage{xcolor}
\usepackage{soul}

\usepackage{hhline}

\usepackage{textfit} 
\usepackage{hyperref}
\begin{document}

\title{Black hole in Nielsen-Olesen vortex}

\author{Kumar J. B. Ghosh}
\email{jb.ghosh@outlook.com}

\affiliation{ University of Denver,\\ Denver, CO, 80210, USA.}
%


\begin{abstract}
In this article, we calculate the classical vortex solution of a spontaneously broken gauge theory interacting with gravity in (2+1)-dimension. We also calculate the conditions for the formation of a (2+1)-dimensional black hole due to magnetic vortex (a Nielsen-Olesen vortex). The semiclassical Hawking temperature for this black hole is calculated, where we see that the temperature of a BTZ black hole increases or decreases without changing the size of the horizon if we insert the magnetic vortex fields in the black hole. Finally, the first law of black hole thermodynamics is described for this particular solution, which shows that the additional work terms from the scalar and gauge fields compensate the change in the temperature relative to its usual value for the BTZ solution. 
\end{abstract}
\maketitle
\newpage
\begingroup
\hypersetup{pdfborder = {0 0 0}}
\endgroup
\newpage

\section{Introduction}\label{sec:introduction}
Vortices or strings are the topological objects in (2+1)-dimensions which arise from some spontaneously broken gauge theories of a classical complex (or two component real) scalar field. These are non perturbative and topologically non trivial solutions of the field equations. One example of these vortices is Nielsen-Olesen vortex \cite{nielsen1973vortex}, where field theory is constructed for the dual string in flat spacetime background. We are trying to extend this concept by considering a curved spacetime background, that is coupling a Nielsen-Olesen vortex to gravity. There are various calculations in (3+1)-dimensions considering strings or monopoles in curved spacetime \cite{lee1992black, dowker1992euclidean, edelstein1993vortices}. \\

In this article we find the behavior of Nielsen-Olesen vortex solutions for the gauge and scalar fields in a curved spacetime background in (2+1)-dimensions. We construct the Hamiltonian formalism of the vortex coupled to gravity and try to compute the corresponding Einstein equations. Without going to some difficult numerical calculations we try to consider a particular case and calculate the form of the corresponding (2+1)-dimensional metric. In the limiting case the vortex becomes a black hole. The famous example of a black hole solution in (2+1)-dimension is BTZ black hole \cite{banados1992black}. In our case the black hole solution having a Nielsen-Olesen vortex with the region outside the horizon described by a BTZ-like solution.\\

In the Semiclassical regime the black holes can radiate \cite{hawking1974black}, so that we can compute the corresponding thermodynamic quantities like temperature and entropy. The famous Bekenstein-Hawking entropy  of a black hole in terms of its horizon area has been derived in various methods, for e.g. \cite{hawking1975particle, bekenstein1972black, bekenstein1973black, bardeen1973four, bekenstein1974generalized, hawking1976black}. Among them we calculate the Hawking temperature adopting the tunneling formalism, \cite{parikh2000hawking, jiang2006hawking, srinivasan1999particle, kerner2006tunnelling, banerjee2008quantum}. Finally an important consequence of these solutions for black hole thermodynamics is discussed, which appears to violate the well known area law.\\

The black hole in (2+1)-dimensional spacetime has been considerably studied including the mass, electric charge, the angular momentum as the only parameters to describe the black hole. In this article we analyze the case by adding a new parameter in the form of magnetic charge to describe the thermodynamic properties of the black hole. The outline of this article as follows. In section \ref{sec:Calculation}, we introduced the Hamiltonian formalism  of the Nielsen-Olesen vortex in non flat background spacetime. Then we calculate the corresponding metric. There after we calculate the semiclassical Hawking temperature if there is a black hole  inside the magnetic vortex. In section \ref{sec:Discussion}, we make some comments on our computation and result.

\section{Calcultion}\label{sec:Calculation}
\subsection {Hamiltonian formalism of the vortex coupled to gravity}
We consider the Einstein-Hilbert action coupled to electromagnetism 
\begin{equation}
S=S_G +S_V, \label{action}
\end{equation}
with
\begin{equation}
S_G = \int d^3 x \sqrt{-^{(3)}g} \left(R-2\Lambda \right)
\end{equation}
is the pure gravity part and
\begin{equation}
S_V = \int d^3 x \sqrt{-^{(3)}g} \left(-\frac{1}{4}F_{\mu\nu}F^{\mu\nu} + \frac{1}{2}D_\mu \phi^a D^\mu\phi^a -V[\phi] \right),
\end{equation}
is the part where gravity coupled to the Maxwell field. Here $^{(3)}g= \text{det}~ g_{\mu\nu}$ ($\mu,\nu = 1, 2, 3$) is the three dimensional metric, $R$ is the Ricci scalar, $\Lambda= -1/l^2$ is the negative cosmological constant, and $\phi^a$ ($a=1, 2$)  is a charged scalar field. The covariant derivative $D_\mu$ in terms of the gauge field $A_\mu$ is defined as
\begin{equation}
D_\mu \phi^a = \partial_\mu \phi^a -e \epsilon^{ab}A_\mu \phi^b,
\end{equation}
and $V[\phi]$ is the symmetry breaking potential:
\begin{equation}
V[\phi] = \frac{\lambda}{4} \left( |\phi|^2 - \eta^2 \right)^2, \label{EQ5}
\end{equation}
with the symmetry breaking coupling constant $\lambda$.\\

We take the field configuration vortex like, that is $\phi$ and $A$ are time independent and the components of the scalar field has the following form:
\begin{equation}
\phi^1 = \eta \chi(r) \cos (n \theta),
\end{equation}
and
\begin{equation}
\phi^2 = \eta \chi(r) \sin (n \theta). \label{Eq7}
\end{equation}
The abelian gauge field 
\begin{equation}
A_\mu = \frac{1}{e}\left( n- P(r) \right) \partial_\mu \theta, \label{Amu}
\end{equation}
where $n$ is a topological quantity called winding number, which takes only integer values.\\

Finite energy condition imply that
\begin{equation}
|\phi|^2 \to\eta^2 ~\text{as}~ r\to \infty,
\end{equation}
from Eq. (\ref{EQ5}) and $A_\mu$ asymptotically be a pure gauge rotation.\\

The above conditions lead us to write down the simplest field configuration for vortex.
\begin{equation}
\chi(0)= 0 ~~~~~\text{and}~ \chi(r)\to 1 ; r\to \infty,  
\end{equation}
where $\eta$ is the vacuum value, and
\begin{equation}
P(0)= n ~~~~~\text{and}~ P(r)\to 0 ; r\to \infty \label{p(n)}
\end{equation}

The action (\ref{action}) can be expressed in the Hamiltonian form as follows,
\begin{equation}
S= \int dt \int_\Sigma d^2x \pi^{ij}\dot{g}_{ij} - N^\perp H_\perp - N^i H_i
\end{equation}
by adding appropriate surface terms, where
\begin{equation}
H_\perp = \frac{1}{\sqrt{g}} \left[ \pi^{ij}\pi_{ij} - (\pi^i _i)^2 \right] - \sqrt{g} \left[ R-2\Lambda^\prime \right] +\frac{1}{2\sqrt{g}} B^2
\end{equation}
is the Hamiltonian constraint with
\begin{equation}
2 \Lambda^\prime = 2\Lambda + \frac{1}{2} g^{ij}(D_i\phi^a)(D_j\phi^a)+ V[\phi]
\end{equation}
and
\begin{equation}
H_i = \pi^j _{i,j}
\end{equation}
are the momentum constraints. $\pi^{ij}$ are the conjugate momenta corresponding to the canonical variables $g_{ij}$, the two dimensional spacial metric, and  $g = \text{det} (g_{ij})$. The magnetic field density $B$ is given by 
\begin{equation}
F_{ij} = \epsilon_{ij} B.
\end{equation}
The lapse function $N^\perp$ and the shift vector $N^i$ appear in the action as Lagrange multipliers; their variation leads to the field equations 
$H_\perp =0$ and $H^i =0$.

\subsection{\label{subsec: Calculation of the metric} Calculation of the metric}
The stationary, axisymmetric (2+1)-dimensional metric can be expressed as the following spacetime line element
\begin{equation}
ds^2 = -N^\perp(r)^2 dt^2 + f^2(r) dr^2 + r^2 (d\theta - N^\theta (r)dt^2),
\end{equation}
With $0\leqslant r < \infty$ and $0\leqslant \theta < 2\pi$. The spacial metric $g_{ij}$ can be written in the following form
\begin{equation}
g_{ij} = \text{diag}\left(f^2(r), r^2\right).
\end{equation}
We can calculate the components of the momentum variable as:
\begin{equation}
\pi^{\theta r} = -\frac{r}{2N^\perp f}(N^\theta)^\prime . \label{pithetar}
\end{equation}
We vary with respect to the shift vector $N^i$, which gives 
\begin{equation}
\pi^{ij} _{,j} = 0 = g^{il} \partial_k \pi^k _l - \frac{1}{2} g^{il} (\partial_l g_{jk}) \pi^{jk} .
\end{equation}
Since $g_{jk}$ is diagonal and $\pi^{jk}$ is off-diagonal, 
\begin{equation}
g^{il} (\partial_l g_{jk}) \pi^{jk} = 0 
\end{equation}
which gives rise to the following result
\begin{equation}
\pi^{\theta r} = \frac{\mathcal{A}}{r^2},\label{pithetar1}
\end{equation}
where $\mathcal{A}$ is a constant determined by the boundary conditions later.
We can rewrite the Hamiltonian constant as
\begin{equation}
H_\perp = f\left[ \frac{2\mathcal{A}^2}{r^3} + \frac{d}{dr}(\frac{1}{f^2})+2\tilde{\Lambda}r + \frac{\tilde{B}^2}{2rf^2}\right] \label{H},
\end{equation}
where
\begin{equation}
2\tilde{\Lambda} = 2\Lambda + \frac{\eta^2 \chi^2 P^2}{r^2}+V[\phi]
\end{equation}
and
\begin{equation}
\tilde{B}^2 = B^2 + \eta^2 r^2 {\chi^\prime}^2.
\end{equation}
Varying with respect to $N^\perp$ gives $H_\perp =0$, which leads to the following equation
\begin{equation}
\frac{d y}{dr} + \frac{\tilde{B}^2}{2r} y = - 2\tilde{\Lambda} r - 2 \frac{\mathcal{A}^2}{r^3}, \label{metric25}
\end{equation}
with the variable $y= \frac{1}{f^2}$.
The equations of motion for the scalar field $(\phi)$ and gauge fields $(A_\mu)$ arising from (\ref{action}) are given by
\begin{equation}
\frac{1}{\sqrt{-g}} \partial_\mu \sqrt{-g} F^{\mu\nu} = -e \epsilon^{ab} \phi^a D^\nu \phi^b
\end{equation}
and
\begin{equation}
\frac{1}{\sqrt{-g}} D_\mu \sqrt{-g} D^\mu \phi^a = \lambda (\phi^2 - \eta^2) \phi^a.
\end{equation}
For the vortex like solution these equations boil down to 
\begin{equation}
\frac{1}{2} \chi^\prime y^\prime + y (\chi^{\prime \prime} + \frac{\chi^\prime}{r}) = \frac{\chi}{r^2} P^2 + \lambda \eta^2 (\chi^2 -1) \chi \label{chi}
\end{equation}
and
\begin{equation}
\frac{1}{2} P^\prime y^\prime + y (P^{\prime \prime} - \frac{P^\prime}{r}) = e  \eta^2 \chi^2 P. \label{P}
\end{equation}
The above equations (\ref{metric25}), (\ref{P}), (\ref{chi}) are highly coupled non linear second order differential equations. This complicated spacetime can admit more than one solution. There could be some nonsingular vortex solutions depending upon the parameters. Instead of solving these differential equations numerically, we can look for simpler solution considering a particular case. we shall go back to the  equation (\ref{H}) and we can rewrite our Hamiltonian constraint as 
\begin{equation}
H_\perp = f \left[ (a^2+b^2)+ c \right]
\end{equation}
where,
\begin{equation}
a= \frac{P^\prime}{\sqrt{2}\sqrt{r}fe}\pm \frac{\eta \chi P}{\sqrt{r}}, 
\end{equation}
\begin{equation}
b= \frac{\eta\chi^\prime \sqrt{r}}{\sqrt{2}f}\pm \frac{\sqrt{\lambda} \sqrt{r} \eta^2}{2}(\chi^2 -1),
\end{equation}
and
\begin{equation}
c= \frac{2\mathcal{A}^2}{r^3}+ \frac{d}{dr}\left( \frac{1}{f^2} \right) + 2\Lambda r \mp \frac{\eta}{\sqrt{2} r f}\left( \sqrt{\lambda} \eta^2 r^2( \chi^2 -1 ) \chi^\prime+ \frac{2P P^\prime \chi}{e} \right).
\end{equation}
For $H_\perp =0 $ we can choose individually $a=0$, $b=0$ and $c=0$ as a possible solution,
which leads to the following equations
\begin{equation}
P^\prime = \mp \sqrt{2} \eta e f P \chi, \label{Pprime}
\end{equation}
\begin{equation}
\chi^\prime = \mp\frac{\sqrt{\lambda}}{\sqrt{2}} \eta f (\chi^2 -1), \label{Chiprime}
\end{equation}
and
\begin{equation}
\frac{d}{dr} y + \mathcal{P}(r) y = \mathcal{Q}(r). \label{f}
\end{equation}
The quantities $\mathcal{P}$ and $\mathcal{Q}$ are given by
\begin{equation}
\mathcal{P} = \pm \frac{\eta}{\sqrt{2}r}\left(\eta r^2 {\chi^\prime}^2 + \frac{{P^\prime}^2}{\eta e} \right), \label{P(r)}
\end{equation}
and
\begin{equation}
\mathcal{Q} = - \left( 2\Lambda r + \frac{2\mathcal{A}^2}{r^3} \right).
\end{equation}
Since the above three equations are coupled differential equations we can solve these in a perturbative approach. In equations (\ref{Pprime}) and(\ref{Chiprime}) we can choose particular form of $f$, say $f_{BTZ}$, to calculate $\chi$  and $P$ and  put the value of these variables in equation (\ref{f}) to obtain the value of the desired metric. For example from a sufficiently large distance from the origin (outside the horizon in case of black hole solution) we can take the following form of ansatz for the variables $\chi$ and $P$.
\begin{equation}
\chi (r) = 1+ \frac{\chi_1}{r} + \frac{\chi_2}{r^2} + ... \label{chi_r}
\end{equation}
and 
\begin{equation}
P (r) = \frac{P_1}{r} - \frac{P_2}{r^2}+ ... \label{p_r}
\end{equation}
and we can impose suitable boundary conditions to know the value of $\chi_i$'s and $P_i$'s (for $i=1, 2, ...$). But now we are mainly focused on the behaviour of the metric (more precisely $f(r)$) not the scalar and gauge fields.\\ 

This differential equation (\ref{f}) has the solution of the following form
\begin{equation}
y = e^{\int \mathcal{P}(r) dr}\int \mathcal{Q}(r) dr  + \mathcal{D}~ e^{\int \mathcal{P}(r) dr}
\end{equation}
where 
\begin{equation}
y= \frac{1}{f^2}.
\end{equation}

We have to vary the action with respect to $g_{ij}$ which is equivalent to vary with respect to `$f$' which leads to the following equation
\begin{equation}
\frac{{N^\perp} ^\prime}{N^\perp} = -\frac{f ^\prime}{f},
\end{equation}
which gives rise to the following relation
\begin{equation}
N^\perp = f^{-1}.
\end{equation}
We substitute the value of $N^\perp$ in (\ref{pithetar}) and we can calculate the value of $N^\theta$ from equation (\ref{pithetar1}) in the following way 
\begin{equation}
(N^\theta)^\prime = -\frac{2\mathcal{A}}{r^3} \Rightarrow N^\theta = C + \frac{\mathcal{A}}{r^2}.
\end{equation}
We have to find three constants $\mathcal{A}$, $\mathcal{D}$ and $C$. We will take the standard results \cite{banados1992black} which is solved for no magnetic field and scalar field. Which gives $\mathcal{D} = -M$, $\mathcal{A}= -\frac{J}{2}$ and $C=0$,  with mass $M$ and angular momentum $J$, which gives rise to the following equations 
\begin{equation}
y = e^{\int_0^r \mathcal{P}(r) dr}\left(\int \mathcal{Q}(r) dr  -M\right).
\end{equation}
The metric under our consideration can be given by 
\begin{equation}
ds^2 = - {N^\perp}^2 dt^2 + {N^\perp} ^{-2} dr^2 + r^2 \left(d\theta +\frac{J}{2r^2} dt \right)^2 \label{metric}
\end{equation}
with
\begin{equation}
{N^\perp}^2 = e^{\int \mathcal{P}(r) dr} \left(-M + \frac{r^2}{l^2}+ \frac{J^2}{4r^2}\right) \label{equation}
\end{equation}
and the negative cosmological constant $\Lambda = - \frac{1}{l^2}$.\\

From the functions $\mathcal{P}(r), \chi (r), P (r)$ described in equations (\ref{P(r)}, \ref{chi_r}, \ref{p_r}) we discuss the asymptotic behavior of the metric (\ref{metric}). At large $r$ the quantities
\begin{equation}
\chi^\prime = - \frac{\chi_1}{r^2} - O(r^{-3}), \text{ and } P^\prime = - \frac{p_1}{r^2} + O(r^{-3}) ,
\end{equation}
which gives
\begin{equation}
\mathcal{P}(r) = \pm \frac{\eta^2 \chi_1^2}{\sqrt{2} r^3} +  O(r^{-4}),
\end{equation}
and 
\begin{equation}
{N^\perp}^2 = e^{\mp \frac{\eta^2 \chi_1^2}{4\sqrt{2} r^4} } \left(-M + \frac{r^2}{l^2}+ \frac{J^2}{4r^2}\right).  \label{eq52}
\end{equation}
For large value of $r$ the quantity  $e^{\mp \frac{\eta^2 \chi_1^2}{4\sqrt{2} r^4} } \to 1$, and Eq. (\ref{metric}) boils down to the usual BTZ black hole metric.

If there is sufficient energy and mass density inside the vortex have it will form a magnetically charged black hole. The positions of the horizons $r_{\pm}$ can be calculated by putting $N^\perp =0$, which gives the following expression
\begin{equation}
r_{\pm} ^2 = \frac{l^2}{2} \left( M \pm \sqrt{M^2 -\frac{J^2}{l^2}} \right). \label{rpm}
\end{equation}
The angular velocity for this black hole can be given by
\begin{equation}
\Omega = \frac{J}{2r_+ ^2}.
\end{equation}

\subsection{Semiclassical Hawking temperature}
The famous semiclassical Hawking temperature has been calculated in the semiclassical approximation in various methodes, for e.g. \cite{hawking1975particle, bekenstein1972black, bekenstein1973black, bardeen1973four, bekenstein1974generalized, hawking1976black}. We would like to compute the Hawking temperature adopting the tunnelling formalism \cite{parikh2000hawking, jiang2006hawking, srinivasan1999particle, kerner2006tunnelling}. A general procedure based on the Hamilton-Jacobi method for calculation of Hawking temperature is done in \cite{banerjee2008quantum}. First the $r$-$t$ sector of the metric is isolated through the following transformation,
\begin{equation}
d \xi = d\theta - \Omega dt. \label{transformation}
\end{equation}
Then 
using the transformation (\ref{transformation}), in the the near horizon approximation, the metric(\ref{metric}) can be written in the form,
\begin{equation}
ds^2 =  - {N^\perp}^2 dt^2 + {N^\perp} ^{-2} dr^2 + r_+ ^2 d\xi^2,
\end{equation}
where the $r$-$t$ sector is isolated from the angular part ($d\xi^2$).
The expression of semiclassical Hawking temperature for $(2+1)$ dimensional black hole is given by
\begin{equation}
T = \frac{\hbar}{4}\left(\text{Im} \int_c \frac{dr}{N_\perp ^2}\right) ^{-1}
\end{equation}
For our case, the expression for $N^\perp$ is given by
\begin{equation}
N^\perp = h(r){N^\perp}_{BTZ}
\end{equation}
with 
\begin{equation}
h(r)= e^{\int_0^r \mathcal{P}(r) dr}
\end{equation}
from equation (\ref{equation}) and $N_{BTZ}$ is the usual lapse function for BTZ black hole.

In our case semiclassical Hawking temperature can be calculated as 
\begin{equation}
T = \frac{\hbar}{4}\left(\text{Im} \int_c l^2 \frac{r^2 dr}{h(r)^2(r^2 - r+^2)(r^2- r_- ^2)}\right) ^{-1}
\end{equation}
with $r_{\pm}$ (\ref{rpm}) is the inner and outer horizon radii of the usual BTZ black hole.\\

Doing the contour integration we can write down the  expression of the Hawking temperature
\begin{equation}
T= h(r_+)T_{BTZ} \label{temperature_vbh}
\end{equation}
where 
\begin{equation}
T_{BTZ} = \frac{\hbar}{2\pi l^2}\frac{r_+^2-r_-^2}{r_+}
\end{equation}
is the Hawking temperature for the usual BTZ black hole and 
\begin{equation}
h(r_+)= e^{\int_0^{r_+} \mathcal{P}(r) dr} \label{hr+}
\end{equation}
with $\mathcal{P}(r)$ is described at (\ref{P(r)}). In principle the factor $\int_0^{r_+} \mathcal{P}(r) dr \neq 0$ that is $h(r_+)\neq 1$ so that the Hawking temperature is not same as ordinary BTZ black hole.

We can also describe the first law of the Black hole thermodynamics
\begin{equation}
dE = T dS + \Omega ~dJ + \phi ~dQ = \frac{\kappa}{8\pi} dA + \Omega ~dJ + \phi ~dQ, \label{firstlaw}
\end{equation}
where $E$ is the energy, $T$ is the temperature of the black hole, $S$ is the entropy, $\kappa$  is the surface gravity, $A$ is the horizon area, $\Omega$  is the angular velocity, $J$ is the angular momentum, $\Phi$  is the electrostatic potential and $Q$ is the electric charge.
If we put $dE =0$ in the above equation (\ref{firstlaw}), we see that the additional work terms from the scalar and gauge fields that try to compensate the change in the temperature relative to its usual value for the BTZ solution.


We shall try to prove the above statement by calculating the thermodynamic quantities explicitly. First, we start with the Bekenstein-Hawking entropy, which is directly proportional to the area of the black hole horizon. From the fundamental postulate of
black hole thermodynamics, the entropy ($S$) \cite{ganai2019report} is defined as
\begin{equation}
S = \frac{A}{4},
\end{equation}
where $A$ is the area of the event horizon. For $(2+1)$-dimensional black hole the area $A= 2 \pi r_+$, so that the the corresponding entropy is calculated as
\begin{equation}
S = \frac{\pi r_+}{2}.
\end{equation}
In our case, the size of event horizon ($r_+$) for BTZ black hole with vortex is equal to the BTZ black hole without the vortex that is 
\begin{equation}
r_+ ^2 = \frac{l^2}{2} \left( M + \sqrt{M^2 -\frac{J^2}{l^2}} \right) = r_{+BTZ} ^2. \label{rp}
\end{equation}
Therefore, the area of the $(2+1)$-dimensional black hole with vortex solution is the same as the area of the usual BTZ black hole, and so as the entropy; that is $S_{vortex} = S_{BTZ}$.

Now, we calculate the other thermodynamic quantities from the solution obtained in the previous subsection (\ref{subsec: Calculation of the metric}).
The metric under our consideration is written as 
\begin{equation}
ds^2 = - \Delta dt^2 + \Delta ^{-1} dr^2 + r^2 \left(d\theta +\frac{J}{2r^2} dt \right)^2 \label{metric1}
\end{equation}
with the lapse function
\begin{equation}
\Delta = e^{\int \mathcal{P}(r) dr} \left(-M + \frac{r^2}{l^2}+ \frac{J^2}{4r^2}\right). \label{equation1}
\end{equation}

In article \cite{ganai2019report, larranaga2008first}, the authors calculated the temperature and other thermodynamic quantities for an usual BTZ-black hole directly from the solution.  Here we shall use the same formula for calculating the same quantities for our solution.

The surface gravity is calculated as:
\begin{eqnarray}
\kappa &=& \frac{1}{2} \frac{\partial \Delta}{\partial r} \mid_{r=r_+} \nonumber \\
&=& \frac{1}{2} \mathcal{P}(r_+) \Delta (r_+) + \frac{1}{2} e^{\int_0^{r_+} \mathcal{P}(r) dr} \left[ \frac{2r_+}{l^2} - \frac{J^2}{8r_+^3} \right]\nonumber \\
&=& e^{\int_0^{r_+} \mathcal{P}(r) dr} \left[ \frac{r_+}{l^2} - \frac{J^2}{16 r_+^3} \right].
\end{eqnarray}
The Hawking temperature $T$ as follows,
\begin{equation}
T = \frac{\kappa}{2\pi}=  e^{\int_0^{r_+} \mathcal{P}(r) dr} \left[ \frac{r}{2\pi l^2} - \frac{J^2}{32 \pi r^3} \right].
\end{equation}
At $r=r_+$
\begin{equation}
\mathcal{P}(r_+) =  \frac{\eta^2 \chi_1^2}{\sqrt{2} r_+^3} +  O\left(r_+^{-4}\right),
\end{equation}
such that the value of the Hawking temperature becomes
\begin{equation}
T = \frac{\kappa}{2\pi}= e^{- \frac{\eta^2 \chi_1^2}{4\sqrt{2} r_+^4} } \left[ \frac{r_+}{2\pi l^2} - \frac{J^2}{32 \pi r_+^3} \right]. \label{hawking_temperature_1}
\end{equation}
The angular velocity $\Omega$ is given by
\begin{eqnarray}
\Omega &=& \frac{\partial \Delta}{\partial J} \mid_{r=r_+} \nonumber \\
&=& e^{- \frac{\eta^2 \chi_1^2}{4\sqrt{2} r_+^4} } \left[ \frac{J}{2r_+^2} \right].
\end{eqnarray}

From the above calculation we show that the entropy $S$ is same as the usual BTZ black hole, i.e. $dS=0$. As a result, there will be changes to the standard thermodynamic parameters, for e.g. $T , ~ \Omega $ etc., to hold the first law of the black hole thermodynamics.

\section{Discussion}\label{sec:Discussion}

We note down some key observations found from our calculation. Firstly the gravity fields are asymptotically BTZ like.
\newline

If we look at the solutions of the horizon, we see that the horizon radius is not changed in our case but the Hawking temperature is changed by the factor $h(r_+)$. So if the vortex like solution appears in a (2+1)-dimensional BTZ black hole it becomes colder or hotter (depending upon the value of $h(r_+)$)without changing the without changing the size of the horizon.  To keep the entropy same as the usual BTZ black hole, the additional work terms from the scalar and gauge fields will compensate the change in the temperature relative to its usual value for the BTZ solution.
\newline

The factor $h(r_+)$ in (\ref{hr+}) depends on the quantity $P(r)$ in (\ref{Amu}) which is related to the winding number $n$ (see equation (\ref{p(n)})). This winding number is a topological quantity which depends on the topology of the field configurations. So the Hawking temperature of the black hole depends on the topology of the field configuration.
\newline

In 2014, Gregory et.al. \cite{gregory2014vortex}
 described the vortex solutions for rotating black holes in (3+1)-dimensions. Although the vortex in (2+1)-dimension is different from the vortex in (3+1)-dimension, because in (3+1)-dimension spacetime, the vortex is a long string in which each of the (2+1)-dimensional slice contains a vortex solution. Also some topological field theories can uniquely be described in odd spacetime dimensions, for e.g. Chern-Simons theory \cite{10.2307/1971013}, which has vast implication in condensed matter theory. This (2+1)-dimension black hole solution with gauge fields may be useful to describe the behavior of Chern-Simons theory in a gravitational background.
\newline

Nowadays gravitational theory is used to describe a holographically dual description of a superconductor AdS/CFT correspondence principle \cite{hartnoll2008building}. Since the above (2+1)-dimensional spacetime (when a magnetic charge coupled to gravity) is asymptotically anti-de Sitter, we can use AdS/CFT correspondence principle to describe some interesting phenomena in (1+1)-dimensional superconductivity.


\bibliography{VBH}   
\end{document}